\documentclass{article}

\usepackage{arxiv}

\usepackage[utf8]{inputenc} % allow utf-8 input
\usepackage[T1]{fontenc}    % use 8-bit T1 fonts
\usepackage{hyperref}       % hyperlinks
\usepackage{url}            % simple URL typesetting
\usepackage{booktabs}       % professional-quality tables
\usepackage{amsfonts}       % blackboard math symbols
\usepackage{nicefrac}       % compact symbols for 1/2, etc.
\usepackage{microtype}      % microtypography
\usepackage{graphicx}

\usepackage{amsmath,amssymb}
\usepackage{xspace}
\usepackage{xcolor}
\usepackage{hyperref}
\usepackage{booktabs}
\usepackage{makecell}
\usepackage[inline,shortlabels]{enumitem}

\usepackage{multirow}

\newcommand{\vect}[1]{\overline{#1}}

\newcommand{\vectIndex}[2]{\ensuremath{{#1}_{#2}}}

\newcommand{\smt}{SMT\xspace}
\newcommand{\omt}{OMT\xspace}

\newcommand{\Ess}{\mathit{Ess}}
\newcommand{\Mon}{\mathit{Mon}}
\newcommand{\ess}{\mathit{ess}}
\newcommand{\any}{\mathit{any}}
\newcommand{\act}{\mathit{act}}
\newcommand{\inh}{\mathit{inh}}
\newcommand{\hard}{\mathit{hard}}
\newcommand{\distinct}{\mathit{distinct}}

\usepackage{array}
\usepackage{hhline}
\newcolumntype{L}[1]{>{\raggedright\let\newline\\\arraybackslash\hspace{0pt}}m{#1}}
\newcolumntype{C}[1]{>{\centering\let\newline\\\arraybackslash\hspace{0pt}}m{#1}}
\newcolumntype{R}[1]{>{\raggedleft\let\newline\\\arraybackslash\hspace{0pt}}m{#1}}

\newtheorem{definition}{Definition}
\newtheorem{example}{Example}

\title{Inference of Qualitative Models from Steady-State Data via Weighted MaxSMT}

\author{
	Ondřej Huvar \\
	Masaryk University, Brno, Czechia \\
	\texttt{xhuvar@fi.muni.cz}
\And
	Nikola Beneš \\
	Masaryk University, Brno, Czechia \\
	\texttt{xbenes3@fi.muni.cz}
\And
	Martin Jonáš \\
	Masaryk University, Brno, Czechia \\
	\texttt{xjonas@fi.muni.cz}
\And
	David Šafránek \\
	Masaryk University, Brno, Czechia \\
	\texttt{xsafran@fi.muni.cz}	
\And
	Samuel Pastva \\
	Masaryk University, Brno, Czechia \\
	\texttt{xpastva@fi.muni.cz}	
}

\begin{document}
\maketitle
\begin{abstract}
Qualitative models provide crucial instruments for modelling complex biological systems. While advances in automated reasoning and symbolic encodings have enabled rigorous inference of these models from data, the process remains highly fragile. First, biological measurement errors inevitably propagate into formal model specifications. Second, when a specification becomes unsatisfiable, distinguishing between fundamental design flaws and minor technical errors is notoriously difficult. This uncertainty often leads to under-specification, as it is unclear which observations are still ``safe'' to incorporate. To overcome these challenges, we introduce a robust inference method based on weighted MaxSMT. By encoding uncertain biological observations as weighted soft constraints, our approach enables the solver to identify a model best reflecting the observations, even with some conflicting constraints. Our method allows for Boolean and multi-valued variable domains, alongside observations derived from discretisation (level constraints) and differential expression (ordering constraints). We show our approach can be used to successfully infer neural cell differentiation models from prior-knowledge networks with approximately 200--1{,}300 genes using ordering constraints on all included genes.
\end{abstract}

\section{Introduction}

Qualitative (logic-based) models of regulatory networks, such as Boolean Networks (BNs)~\cite{kauffman1969homeostasis} or Thomas Networks (TNs)~\cite{thomas1991regulatory}, represent an abstract yet mechanistic framework for studying the non-linear dynamics emerging from influences (regulations) among the system's components. Qualitative models have proven useful for modelling many processes in living cells, including gene regulation and protein activities in signal transduction. Every qualitative model consists of variables, each associated with an update function governing its dynamics based on the current state and the effect of incoming regulations. The goal of qualitative model inference is to reconstruct a particular model from experimental observations and other prior knowledge. Inference of models that reproduce experimental observations, satisfy desired structural and dynamic properties, and reflect the expert knowledge remains a challenge.

%This becomes even more critical when taking into account single-cell transcriptomics data.

% something about partially specified models, prior knowledge, influence graph, different kinds of data, focus on _steady state observations_

%Existing inference methods for qualitative models mostly focus on Boolean Networks and require the data to be preprocessed by a suitable binarisation (discretisation) technique. They can be broadly divided into two classes: optimization-based approaches, which aim to identify one or several best-scoring candidate models, and exhaustive approaches, which characterize the full set of models consistent with the available data and prior knowledge. The former class often relies on heuristic or score-driven procedures, including genetic programming~\cite{gao2020learning}, Best-Fit extensions~\cite{lahdesmaki2003learning}, or mutual-information-based methods~\cite{barman2017novel}, and therefore typically does not provide strong guarantees of completeness with respect to the space of admissible solutions. Exhaustive methods~\cite{chevalier2024bonesis,benevs2023boolean,yordanov2016method,yordanov2023reasoning}, in contrast, return exact results, but they face the fundamental challenge of combinatorial explosion of the candidate set, which often renders full enumeration infeasible in practice. In general, the outcome of most methods in both families strongly depends on the settings of the data discretisation procedure. Moreover, the methods are typically tightly coupled with the particular update scheme driving the model dynamics.
Existing inference methods for qualitative models mostly focus on BNs and require the data to be preprocessed by a suitable binarisation (discretisation) technique.
These inference methods can be broadly divided into two classes.
The first consists of heuristic optimisation-based approaches that employ score-driven procedures to identify one or several high-scoring candidate models, without strong guarantees on completeness or optimality.
These approaches, recently reviewed in~\cite{Pusnik2022Review}, include methods based on genetic programming~\cite{gao2020learning}, best-fit extensions~\cite{lahdesmaki2003learning}, or mutual information~\cite{barman2017novel}.
The second class consists of exact approaches that utilise formal methods~\cite{chevalier2024bonesis,benevs2023boolean,yordanov2016method,yordanov2023reasoning} to compute candidate models that are guaranteed to be consistent with the model specification.
However, the current exact methods are poorly suited for noisy experimental data and critically suffer from scalability issues.
In general, the outcome of most methods in both families strongly depends on the settings of the data discretisation procedure.

In~\cite{huvar2026smtinference}, we tackled the inference problem of qualitative models by using Satisfiability Modulo Theories (SMT). The unknown update functions are encoded directly as uninterpreted functions subject to additional constraints derived from fixed-point observations and the prior knowledge of the influence graph.

In this paper, we utilise the optimisation techniques developed within SMT (specifically, weighted MaxSMT~\cite{bjorner2015muz,sebastiani2020optimathsat}) to overcome the problems of existing inference methods mentioned above. Our inference problem is set up for Multi-Valued Networks (MVNs) where each variable has a finite discrete domain~\cite{ginsim,schaub2007qualitative} (i.e., MVNs cover TNs as well as BNs). The input to the inference procedure is: (i) the influence graph associated with ``regulation constraints'' (monotonicity and essentiality) expressing prior knowledge about the effects of specific influences; (ii) the ``observation constraints'', asserting the existence of several data-informed fixed-point states (biological steady states). Although BN and MVN models admit several alternative semantics representing their concurrent dynamics, the concrete choice of the update scheme becomes immaterial when the focus is restricted to steady states. Consequently, our approach is semantics-agnostic.

Constraints of our inference problem can be prescribed as \emph{hard} (required) or \emph{soft} (not necessarily required to be satisfied). Hard constraints must be satisfied by every admissible model. Soft constraints are associated with weights and provide a natural mechanism for representing uncertainty in biological knowledge and observations. The weights reflect the strength (or reliability) of the corresponding experimental evidence, and the objective is to identify candidate models that maximise the total weight of the satisfied soft constraints. Hence, unlike purely satisfiability-based formulations, we do not require all observations to be jointly satisfiable. Instead, we compute a model that best explains the imprecise steady-state observations, while also adhering to all prescribed hard regulation constraints.

Our contribution proceeds as follows. First, we introduce the constraint-based inference problem for MVNs. Second, we provide the SMT encoding of the inference problem, defining the regulation and observation constraints, including their weighting and the objective function construction. Third, we describe the workflow for translating prior knowledge and experimental observations into a set of weighted constraints. Finally, we demonstrate the applicability of the method on an extensive set of inference problem instances targeting differentiation of neural tissue cells in the mouse cerebral cortex using an scRNA-seq expression dataset. Influence graphs of the inferred MVNs represent large gene regulatory networks (from 200 to $\geq 1{,}300$ genes) explaining the processes behind cell differentiation. Steady states correspond to phenotypes representing individual cell types. The results are supported with a reproducible artefact.

\paragraph{Related Work.}
To the best of our knowledge, the existing tools for exact inference do not employ optimisation to overcome uncertainty in data. Moreover, most of the existing tools focus on Boolean models only. The only other line of methods we are aware of that employs SMT encoding of the inference problem is based on
RE:IN~\cite{yordanov2016method,yordanov2023reasoning}. It utilises satisfiability without optimisation. The constraints are syntactically restricted to a limited subset of update functions \emph{without} uninterpreted functions (employing a finite library of biologically motivated regulation conditions). In~\cite{yordanov2022smt}, SMT-based frameworks such as REIL generalise this line of work to broader classes of partially specified discrete dynamical systems with finite-domain variables. BoNesis~\cite{chevalier2024bonesis,chevalier2025data} and
Caspo~\cite{guziolowski2013exhaustively} rely on Answer Set Programming (ASP) with exhaustive Disjunctive Normal Form (DNF)
encodings. Finally,
AEON~\cite{benevs2022aeon,benevs2023boolean} presents a symbolic approach employing Binary Decision Diagrams (BDDs) to exhaustively represent the
update function space. %The BDD encoding---for each update function and for each combination of its input values---introduces one BDD variable representing the result of the function. This induces an exponential blowup, {\color{red} restricting the tool's applicability to functions of low arity. -- D: but this problem is not completely avoided with SMT in general, only in cases of a limited set of contraints, right?}

%Our method is not restricted to Boolean models, and it accounts for the fact that experimental observations are typically uncertain and mutually inconsistent. This makes it possible to capture the unavoidable noise and partial inconsistency present in experimental data and literature knowledge, which is not directly supported by the above-mentioned methods.

\enlargethispage*{4mm}
\section{Preliminaries}

We first present the formalism of multi-valued networks.
We then provide a brief overview of satisfiability modulo theories and their optimisation extension.

\subsection{Multi-Valued Networks}

\begin{definition}[Multi-valued network]
	\label{def:mvn}
	A \emph{multi-valued network (MVN)} is a triple $\mathcal{M} = (V, D, F)$, such that
	\begin{itemize}
		\item $V = \{ v_1, \dots, v_n \}$ is an indexed set of \emph{network variables}.
		\item $D = \{ D_1, \dots, D_n \}$ is an indexed set of \emph{variable domains}; each $D_{i}$ is either Boolean ($D_{i} = \{ 0, 1 \}$) or a finite integer interval (e.g., $D_{i} = \{ 0, \ldots, 5 \}$).
		\item $F = \{ F_1, \dots, F_n \}$ is an indexed set of \emph{update functions}. Each variable $v_i$ is equipped with an \emph{update function} $F_{i} \colon \prod_{j=1}^n D_{j} \rightarrow D_{i}$.
	\end{itemize}
	If all the variable domains are Boolean, $\mathcal{M}$ is called a \emph{Boolean network} (BN).
\end{definition}

A \emph{system state} $x \in \prod_{j=1}^n D_{j}$ is a~mapping that assigns each network variable $v_i$ a value $x[i] \in D_{i}$.
A state $x$ is a~\emph{fixed-point state} if $F_{i}(x) = x[i]$ for all $1 \leq i \leq n$.
%A fixed-point state $x$ is a stable configuration of the MVN such %that for each $v_i$, we have $F_{i}(x) = x[i]$.
%The behaviour of the system is governed by the update functions, which specify how each variable evolves based on the system's current state. The particular state-transition graph also depends on a selected update scheme, which determines how individual variable changes combine to make a transition. \red{todo: general vs unitary}. \red{todo: update concurrency}. However, the set of fixed-point states is preserved across all standard update schemes.

Each update function $F_{i}$ typically depends only on a subset of variables $dep(F_{i}) \subseteq V$. These variables $v_j \in dep(F_{i})$ are called \emph{essential inputs of $F_i$} or \emph{regulators of $v_i$}.
Furthermore, the regulators may exhibit different \emph{monotonicities}. A regulator $v_j$ is called an \emph{activator (inhibitor)} of $v_i$ if $F_{i}$ is positively (resp. negatively) monotone with respect to $v_j$. Some regulators may also have non-monotone effects, i.e., they are neither positively nor negatively monotone.

\subsection{Satisfiability Modulo Theories and Optimisation}

Given a first-order formula $\varphi$ over a given logical theory $\mathcal{T}$, the goal of \emph{Satisfiability Modulo Theories} (\smt)~\cite{barrett2021smt} is to decide whether there is a model of the theory $\mathcal{T}$ that satisfies $\varphi$. \smt solvers implement efficient decision procedures for various theories. In this paper, we consider the theory of \emph{linear integer arithmetic with uninterpreted functions} ($\mathsf{UFLIA}$), where the formulas consist of integer variables, integer constants (such as $2$ or $-42$), the addition function $+$, and the relational operators ${\bowtie} \in \{ =, \neq, \leq, \geq, <, >\}$. Additionally, formulas can contain \emph{uninterpreted function symbols} (such as $f$, $g$, $\ldots$). The model of such a formula $\varphi$ over $\mathsf{UFLIA}$ is an assignment of integers to the free variables in $\varphi$ and of functions to the uninterpreted function symbols such that the assignment satisfies $\varphi$.

The \smt problem can be extended with optimisation, resulting in a problem known as \emph{Optimisation Modulo Theories} (\omt)~\cite{sebastiani2020optimathsat}. In \omt, the problem is for the given formula $\varphi$ over the theory $\mathcal{T}$ and an objective function $\mathit{obj}$ to find a model $\mu$ of $\varphi$ with a maximal (or dually, minimal) value of $\mathit{obj}(\mu)$.
The technique can be extended to multiple objective functions $\mathit{obj}_1$, $\ldots$, $\mathit{obj}_k$; the goal is then to find a model that \emph{lexicographically} maximises the value of the tuple $(\mathit{obj}_1(\mu), \mathit{obj}_2(\mu), \ldots, \mathit{obj}_k(\mu))$.

A special case of \omt, which is supported by several existing solvers~\cite{bjorner2015muz,sebastiani2020optimathsat}, is \emph{weighted MaxSMT}. In weighted MaxSMT, the input is a set of formulas $C_h$, called \emph{hard constraints}, and a set of formulas $C_s$, called \emph{soft constraints}. Each soft constraint $c \in C_s$ has an assigned non-negative \emph{weight} $w(c)$. The goal of weighted MaxSMT is then to find an assignment $\mu$ that \emph{satisfies all the hard constraints} and that maximises the sum of weights of the satisfied soft constraints $\mathit{score}(C_s, \mu) = \sum \{ w(c) \mid c \in C_s, \mu \text{ satisfies } c \}$.
As an example, consider the hard constraints $C_h = \{ x + y = z,~ y \geq 10 \}$ and soft constraints $C_s = \{ x \geq 0,\, z \leq 0,\, x + z \geq 40 \}$ with weights $8$, $3$, and $6$, respectively. An optimal solution is $\mu(x) = 20$, $\mu(y) = 10$, $\mu(z) = 30$, which has score $8+6=14$.
The problem can be naturally extended to several priority classes $C_s^1$, $C_s^2$, $\ldots$, $C_s^k$ of soft constraints. The goal is then to maximise the value of the tuple $(\mathit{score}(C_s^1, \mu), \mathit{score}(C_s^2, \mu), \ldots, \mathit{score}(C_s^k, \mu))$ lexicographically. Intuitively, hard constraints must be satisfied, and the score achieved by satisfied constraints from $C_s^i$ is more important than the score of each $C_s^j$ with $j > i$.

Notably, most of the existing solvers for weighted MaxSMT implement \emph{anytime algorithms}, i.e., during the computation, they produce solutions with increasingly better scores. If they are stopped early, they still output a valid solution to the hard constraints, albeit possibly not a maximal one.

\section{Weighted MVN Inference Problem}
\label{sec:mvn-inference}

This section defines the main problem addressed in our paper: \emph{the weighted
	MVN inference
	problem}. %and introduces our approach that relies on MaxSMT to solve the inference problem. %and how to encode it into SMT and derive an optimal model with respect to the specified constraints.
Its input is the \emph{inference specification} consisting of
\begin{enumerate*}[(a)]
	\item regulation constraints derived from a prior-knowledge influence
	graph (incorporating monotonicity and essentiality) that restrict the structure of the inferred model, and
	\item fixed-point observation constraints that restrict the model's dynamics.
\end{enumerate*}
The fixed-point specification can come from noisy
or otherwise uncertain observations, and can thus be treated either as hard or as soft constraints
with assigned weights, depending on our confidence in these observations. The goal is to find an MVN that satisfies all the hard constraints and maximises the sum
of the weights of the satisfied observations.

\paragraph{Regulation constraints}
The general structure of the model is determined by a directed
\textit{influence graph} $\mathcal{I}=(V, E)$, where nodes
$V = \{v_1, \ldots, v_n\}$ are the network variables and edges
$E \subseteq V \times V$ represent potential \emph{regulations} (i.e.,
$v_i$ can regulate $v_j$ only if $(v_i, v_j) \in E$). For each variable
$v_i \in V$, we prescribe an associated domain $D_{i}$ (Boolean or
multi-valued). Finally, some of the regulations are constrained to be essential or have a specific monotonicity.

A regulation $(v_i, v_j) \in E$ is \emph{essential} if the regulator $v_i$ has an observable influence on the target $v_j$.
Regulations without an essentiality constraint are optional and are not required to influence the target's update function at all. A regulation can also be assigned a \emph{monotonicity}, forcing the regulator to act as either an activator or an inhibitor.

The \emph{regulation constraints} are prescribed by a pair of functions $(\Ess, \Mon)$ with signatures $\Ess \colon E \to \{\ess, \any\}$ ($\ess$ for essential regulations, $\any$ for unspecified essentiality) and $\Mon \colon E \to \{\act, \inh, \any\}$ ($\act$ for activations, $\inh$ for inhibitions, and $\any$ for unspecified monotonicity).

\begin{definition}[Regulation constraints consistency]
	\label{def:regulation-consistency}
	Let $\mathcal{I} = (V, E)$ be an~influence graph and $\mathcal{M} = (V, D, F)$ an MVN over the same $n$ variables. An update function $F_i \in F$ is \emph{consistent} with $\mathcal{I}$ if $dep(F_i) \subseteq \{ v_j \mid (v_j, v_i) \in E \}$.
	
	Furthermore, let $(\Ess, \Mon)$ be regulation constraints on $E$. A~function $F_i \in F$ is consistent with $(\Ess, \Mon)$ if all $(v_j, v_i) \in E$ satisfy:
	\vspace{-1mm}
	\begin{itemize}
		\item if $\Ess(v_j, v_i) = \ess$, then $v_j \in dep(F_i)$,
		\item if $\Mon(v_j, v_i) = \act$, then $F_i$ is positively monotone with respect to $v_j$, and
		\item if $\Mon(v_j, v_i) = \inh$, then $F_i$ is negatively monotone with respect to $v_j$.
	\end{itemize}
\end{definition}

%Second, the dynamics of the model are constrained in a following way.
\paragraph{Observation constraints}
Moreover, the model is required to exhibit the prescribed \emph{fixed points}, with specific \emph{observation constraints} placed on the values of network variables in these states.
Each of the $m$ required fixed-point states is specified as a vector $s^j$ of \emph{state variables} $(s_1^j, \ldots, s_n^j)$, where $s_i^j$ represents the value of $v_i$ in the fixed-point state~$s^j$.
An additional Boolean parameter $\distinct \in \{0, 1\}$ specifies whether the fixed-point states are constrained to be pairwise distinct.

The admissible values of the fixed-point state variables are restricted by a set of observation constraints $\mathcal{C}$. Each $c \in \mathcal{C}$ is defined by a relational operator ${\bowtie} \in \{=, \neq, \geq, \leq, >, <\}$ comparing either two state variables ($s_i^j \bowtie s_{i'}^{j'}$), or a state variable and an integer constant ($s_i^j \bowtie k$).
Each $c \in \mathcal{C}$ has a weight assigned by a \emph{weighting function} $w \colon \mathcal{C} \to {\mathbb{Q} \cup \{\hard\}}$.
All $c$ with $w(c) = \hard$ are \emph{hard} constraints that must always be satisfied, while the others are considered \emph{soft}.

Since observation constraints may qualitatively differ in their importance, they are divided into $l+1$ disjoint categories (levels) $\{\mathcal{C}_{\hard}, \mathcal{C}_1, \ldots, \mathcal{C}_l\}$, creating a partition of~$\mathcal{C}$.
While $\mathcal{C}_{\hard}$ represents the hard constraints, the remaining sets $\mathcal{C}_1, \ldots, \mathcal{C}_l$ contain soft constraints whose qualitative importance decreases with larger $l$. The motivation for such partitioning is explained in Section~\ref{sec:constraints}.

\begin{figure}[t]
	\begin{minipage}{0.4\linewidth}
		\centering
		\includegraphics[width=0.7\linewidth]{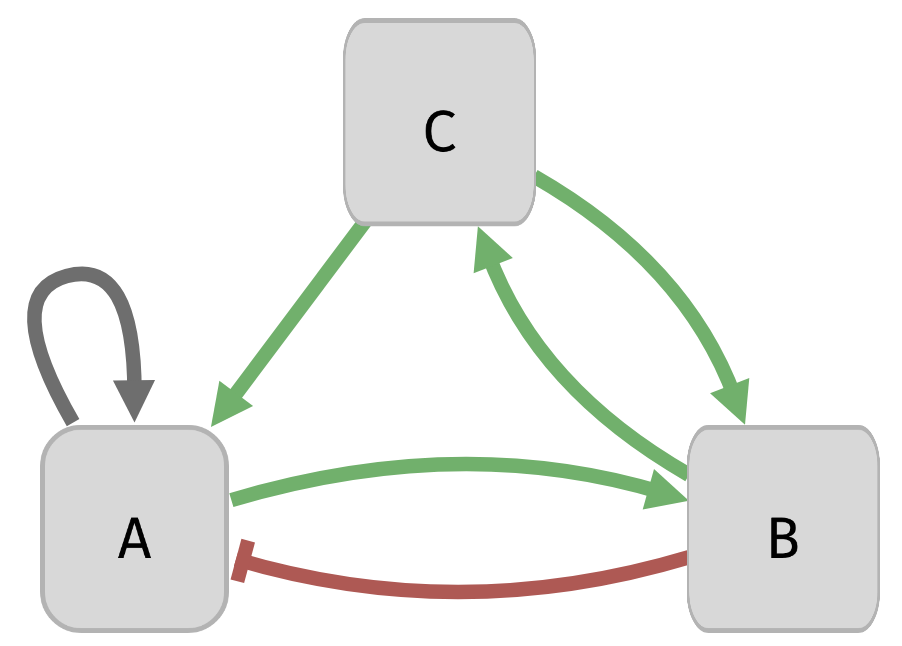}
		\begin{align*}
			D_A &= \{0, 1\} \\
			D_B &= \{0, 1, 2\} \\
			D_C &= \{0, 1, 2\}
		\end{align*}
		(a) Influence graph
	\end{minipage}%
	\begin{minipage}{0.6\linewidth}
		\centering
		\textbf{Scenario I}\\
		Fixed points: $(s^1_A, s^1_B, s^1_C)$, $(s^2_A, s^2_B, s^2_C)$
		\setlength{\tabcolsep}{4pt}
		\begin{tabular}{ccccccc}
			\hline
			value & $s^1_A$ & $s^1_B$ & $s^1_C$ & $s^2_A$ & $s^2_B$ & $s^2_C$
			\\ \hline
			0 & \textbf{0.74} & 0.07 & \textbf{0.53} & 0.07 & \textbf{0.57} & 0.20
			\\
			1 & 0.26 & 0.11 & 0.28 & \textbf{0.93} & 0.22 & \textbf{0.61}
			\\
			2 & --- & \textbf{0.82} & 0.19 & --- & 0.21 & 0.19
			\\ \hline
		\end{tabular}
		
		\smallskip
		(b) Inference specification I
		
		\medskip
		\textbf{Scenario II}\\
		Fixed points: $(s^1_A, s^1_B, s^1_C)$, $(s^2_A, s^2_B, s^2_C)$, $(s^3_A, s^3_B, s^3_C)$
		\setlength{\tabcolsep}{5pt}
		\begin{tabular}{ccccc}
			\hline
			$\bowtie$
			& $s^1_B \bowtie s^2_B$
			& $s^1_C \bowtie s^2_C$
			& $s^2_B \bowtie s^3_B$
			& $s^2_C \bowtie s^3_C$
			\\\hline
			$<$ & 0.05 & \textbf{0.52} & 0.04 & \textbf{0.45}
			\\
			$=$ & 0.34 & 0.40 & 0.31 & 0.44
			\\
			$>$ & \textbf{0.61} & 0.08 & \textbf{0.65} & 0.11
			\\\hline
		\end{tabular}
		
		\smallskip
		Hard constraints: $s^1_A = s^2_A = s^3_A = 1$
		
		\smallskip
		(c) Inference specification II
	\end{minipage}
	\caption{%
		(a) An influence graph of a three-variable MVN with regulation constraints. The grey arrow (the self-loop over $A$) denotes a regulation with no assigned monotonicity; the red arrow (from $B$ to $A$) denotes an inhibition; the remaining green arrows represent activations. All regulations are essential. The monotonicity and essentiality constraints have to be satisfied by all feasible solutions.
		(b) An example of observation constraints in the form of equality with constants (e.g., discretised activity levels). There are two fixed points, with 16 soft constraints succinctly represented in a table together with their weights: $s_A^1 = 0$ with weight 0.74, $s_B^1 = 1$ with weight 0.11, etc. The maximal weights are depicted in boldface.
		(c) An example of observation constraints formulated as a comparison between fixed-point state values (e.g., differential expression). There are three fixed points, with the hard constraints that the value of $A$ is 1 in all of them.
		There are 12 soft constraints represented in the table: $s_B^1 < s_B^2$ with weight 0.05, etc.%
	}%
	\label{fig:example}
\end{figure}

\begin{example}\label{ex:running}
	Two example scenarios (I and II) of the inference specification are given in Fig.~\ref{fig:example}, both sharing the same influence graph and domains (Fig.~\ref{fig:example}a).
	For better clarity, we use the names $A$, $B$, $C$ for variables instead of $v_1$, $v_2$, $v_3$.
	The influence graph has six regulations, all of which are essential. The monotonicity of the self-regulation on $A$ is unconstrained, the regulation $(B, A)$ is an inhibition, and all the remaining regulations are activations. The domain of $A$ is Boolean, while the domains of $B$ and $C$ are three-valued.
	
	In Scenario I (Fig.~\ref{fig:example}b), the specification requires two fixed points. Here, every constraint is given as an equality with an integer, and the table shows the constraints' weights, i.e., $s^1_A = 0$ with the weight of 0.74, etc. For simplicity, the constraints are all assigned the same level of importance.
	
	In Scenario II (Fig.~\ref{fig:example}c), the specification requires three fixed points. Three hard constraints state that the value of $A$ is 1 in each fixed point. Soft constraints are equalities and inequalities over pairs of state variables, i.e., $s^1_B < s^2_B$ with the weight of 0.05, $s^1_B = s^2_B$ with the weight of 0.34, etc. Again, all the soft constraints are given the same level of importance.
\end{example}

%Altogether, the constraints are divided into two categories. The \textit{hard} constraints consist of required regulation properties (monotonicity and essentiality), and the existence of required unique fixed-point states.
%The \textit{soft} constraints then describe variable assignments in particular fixed-point states and relations between them. Each of these soft constraints is given a weight, as discussed in Section~\ref{subsec-optimization}.

\begin{definition}[Weighted MVN Inference Problem]
	\label{def:inference}
	Let $n$ be the number of variables and $m$ the number of fixed points. Additionally:
	\begin{itemize}
		\item Let $D = \{ D_1, \ldots, D_n \}$ be an indexed set of variable domains;
		\item Let $\mathcal{I} = (V, E)$ be a network influence graph over $n$ variables;
		\item Let $(\Ess, \Mon)$ be the regulation constraints on $E$;
		\item Let $S = \{s_i^j \mid 1 \leq i \leq n, 1 \leq j \leq m\}$ be the state variables of the $m$ fixed points, with $\distinct \in \{0, 1\}$ being the distinctness parameter;
		\item Let $\mathcal{C} = \mathcal{C}_{hard} \cup \mathcal{C}_1 \cup \ldots \cup \mathcal{C}_l$ be a set of observation constraints partitioned into $l+1$ levels, with $w$ as their weighting function.
	\end{itemize}
	
	% Let $\mathcal{I} = (V, E)$ be a network influence graph over $n$ variables, $D$ be an indexed set of variable domains, and $(\Ess, \Mon)$ regulation constraints on $E$.
	% Let $S = \{s_i^j \mid 1 \leq i \leq n, 1 \leq j \leq m\}$ be a set of state variables representing $m$ fixed points, $\distinct \in \{0, 1\}$ a fixed-point distinctness parameter, $\mathcal{C} = \mathcal{C}_{hard} \cup \mathcal{C}_1 \cup \ldots \cup \mathcal{C}_l$ a set of observation constraints partitioned into $l+1$ levels, and $w$ a~constraint weighting function.
	
	A \emph{feasible solution} of the MVN inference problem is a pair $(\mathcal{F}, \mathcal{S})$ consisting of an indexed set of functions $\mathcal{F} = \{F_{1}, \ldots, F_{n}\}$ and a function $\mathcal{S}$ mapping each $s_i^j$ to a value from $D_i$, such that:
	\vspace{-1mm}
	\begin{itemize}
		\item $\mathcal{M} = (V, D, \mathcal{F})$ is a valid multi-valued network,
		\item each function $F_i \in \mathcal{F}$ is consistent with $\mathcal{I}$ and with constraints $(\Ess, \Mon)$,
		\item for each $1 \leq j \leq m$, $(\mathcal{S}(s_1^j), \ldots, \mathcal{S}(s_n^j))$ is a fixed-point state of~$\mathcal{M}$, all of which must be pairwise distinct if $\distinct=1$, and
		\item all constraints $c \in \mathcal{C}_{hard}$ are satisfied by the assignment $\mathcal{S}$.
	\end{itemize}
	
	We write $\mathcal S \models c$ to denote that the assignment
	$\mathcal S$ satisfies the constraint $c \in \mathcal{C}$.
	For each level $k$, the $k$-level weight of a feasible solution
	is given as
	\[w_k(\mathcal S) = \sum \bigl\{w(c) \mid c \in \mathcal C_k, \mathcal S \models c\bigr\}.\]
	%where $i_{\mathcal{S}, c} \in \{0, 1\}$ indicates whether the constraint $c$ is satisfied by the assignment $\mathcal{S}$ ($i_{\mathcal{S}, c} = 1$) or not ($i_{\mathcal{S}, c} = 0$).
	The \emph{weight} of a feasible solution is then defined as an $l$-tuple $(w_1(\mathcal S), \ldots, w_l(\mathcal S))$.
	A feasible solution is \emph{optimal} if its weight is a lexicographic maximum of the weights of all feasible solutions.
	
\end{definition}

\begin{example}
	We now continue with the examples from Fig.~\ref{fig:example}.
	In Scenario I, there are no feasible solutions that satisfy all the maximally-weighted constraints (i.e., constraints with highest user confidence; depicted by boldface in Fig.~\ref{fig:example}b), since having (0, 2, 0)
	as a fixed point is impossible due to the essentiality and monotonicity constraints.
	An optimal solution instead assigns the values (0, 2, 1) and (1, 2, 1) to the fixed-point states.
	The weight of such solution is 3.59. The corresponding update functions are
	described in Appendix~\ref{app:example} of~\cite{huvar2026inference}. Note that although the fixed point (1, 0, 1) would also be allowed by the essentiality and monotonicity requirements, it cannot appear together with a fixed point (0, 2, 1), and all other solutions including (1, 0, 1) have a worse weight.
	Also note that there are, in fact, many optimal solutions (with different choices of update functions), but they all agree on the two fixed-point states listed here.
	
	In Scenario II, there are again no feasible solutions that satisfy all the maximally-weighted constraints. If we do not constrain the fixed points to be pairwise distinct, there are many optimal solutions leading to a variety of fixed-point states. However, all share the property that the first and the second fixed point are the same. One such possibility is
	(1, 2, 2), (1, 2, 2), and (1, 1, 1), with the weight of 1.50. Adding the distinctness constraint, we get an optimal solution with the fixed points
	(1, 2, 2), (1, 1, 1), and (1, 0, 0) and the weight of 1.45. This suggests an issue with the observations on which the constraints are based. For example, if each fixed point is meant to represent a dedicated cell type, the best explanation of the data that the model can provide instead suggests that the first and second cell type are in fact the same.
	
	% Such a situation, where the optimal solutions are only those where
	% some of the fixed-point states are not distinct, can be interpreted as an issue with the data on which the user based the constraints. One possible interpretation, in the case where the fixed-point states are supposed to represent different cell types, is that the first and the second cell types are in fact the same.
\end{example}

\section{Encoding Weighted MVN Inference in MaxSMT}
\label{sec:mvn-inference-encoding}

% Influence graph, uninterpreted functions, fixed-points, monotonicity, essentiality, uniqueness of fixed-points...

%The SMT query to encode the model specification is constructed in the following way.

We now show that the problem of weighted MVN inference
naturally maps to MaxSMT over linear integer arithmetic with
uninterpreted functions. To encode the model specification, we
construct SMT formulas for regulation monotonicity and fixed-point
constraints by the approach introduced
in our previous work~\cite{huvar2026smtinference}. Subsequently, we add the observation
constraints with the desired weights. The optimal models of the
resulting formula directly correspond to solutions to the inference
problem.

First, an uninterpreted function $f_{i}$ is declared for each $v_i$ to
represent its update function. The arity of each $f_{i}$ corresponds
to the in-degree of $v_i$ in the influence graph (only the variable's
regulators can influence the output of its update function), and the
function domain matches the regulators. In addition, all state variables
$s_i^j$ are treated as variables in the SMT formulas.
%a fresh constant $\varsigma_i^j$ is declared to represent each state variable $s_i^j$.

The following SMT formulas are then constructed and asserted to impose
constraints on the uninterpreted functions and state variables, obtaining an interpretation that
%is consistent with the inference specification while maximizing the objective.
corresponds to an optimal solution to the MVN inference.

\smallskip
\noindent
\emph{Regulation essentiality.}
%Some regulations of the influence graph may be declared \emph{essential}, which means that the regulator variable must have an observable effect on the target.
A requirement of regulation $(v_i, v_j) \in E$ being essential naturally translates to a constraint requiring that the input of the function $f_{j}$ corresponding to $v_i$ is essential. Assuming without loss of generality that $v_i$ is the first of $k$ regulators for $v_j$, we encode the essentiality as the hard constraint $\eta_{i, j}$ defined in Equation~\ref{eq:essentiality}.

\begin{equation}
	\label{eq:essentiality}
	\eta_{i, j} \equiv \exists a, b, e_2, \ldots, e_k .\, f_{j}(a, e_2, \ldots, e_k) \neq f_{j}(b,e_2, \ldots, e_k)
\end{equation}

\smallskip
\noindent
\emph{Regulation monotonicity.} The monotonicity of a regulation $(v_i, v_j) \in E$ also naturally translates into a hard constraint requiring the function $f_{j}$ to be monotone with respect to its input corresponding to $v_i$.
%We use an approach proposed in~\cite{sofronie2007automated,sofronie2014ieee} to encode the monotonicity of a function with respect to a subset of its arguments. For any function $g$, we denote the set of all indices of its positively monotone inputs as $\mathcal{P}_g$ and the set of indices of negatively monotone inputs as $\mathcal{N}_g$. These two sets can be easily derived from the regulation monotonicity specification $Mon$. Given a function $g$ (of arity $k$) and the two sets $\mathcal{P}_g$ and $\mathcal{N}_g$ specifying its monotone inputs, the monotonicity of $g$ is enforced via the formula template $\psi_g$ defined in Equation~\ref{eq:monotonicity-aggregated}.
Following the approach proposed in~\cite{sofronie2007automated} and evaluated in~\cite{huvar2026smtinference}, the monotonicity of any function $g$ (of arity $k$) can be enforced via the formula template $\psi_g$ defined in Equation~\ref{eq:monotonicity-aggregated}.
Here, $\mathcal{P}_g$ and $\mathcal{N}_g$ denote the sets of positively and negatively monotone input indices, respectively, while $\mathcal R_g$ denotes the set of remaining (non-monotone) input indices. These sets can be easily derived from the monotonicity specification $\mathit{Mon}$.
We use $\overline{x}$ as a shorthand for $x_1, \ldots, x_k$ (same for $\overline{y}$). Using this template, the monotonicity of each update function $f_{i}$ can be ensured by constructing its corresponding $\psi_{f_i}$. Furthermore, to avoid universal quantifiers in each $\psi_{f_i}$, we use the quantifier instantiation approach introduced in~\cite{sofronie2007automated}, which, as shown in~\cite{huvar2026smtinference}, significantly improves solver performance on practical problem instances.

{
	\small % Otherwise the equation won't fit on one line.
	\begin{equation}
		\label{eq:monotonicity-aggregated}
		\forall \overline{x}, \overline{y}.
		\left(
		\left ( \bigwedge_{\substack{i \in \mathcal{P}_g}} \vectIndex{x}{i} \leq \vectIndex{y}{i} \right ) \wedge
		\left ( \bigwedge_{\substack{i \in \mathcal{N}_g}} \vectIndex{y}{i} \leq \vectIndex{x}{i} \right ) \wedge
		\left ( \bigwedge_{\substack{i \in \mathcal{R}_g}}{\vectIndex{x}{i} = \vectIndex{y}{i}} \right )
		\right )
		\Rightarrow
		g(\vect{x}) \leq g(\vect{y})
	\end{equation}
}

\smallskip
\noindent
\emph{Fixed-point states.}
For each fixed-point state $s$ in the specification, we enforce that $(s_1, \ldots, s_n)$ is indeed a fixed point by asserting the hard-constraint formula $\sigma_s$ constructed according to the template in Equation~\ref{eq:fixed-point-macro}, where $r^i_1, \ldots, r^i_k$ denote the indices of the $k$ regulators of $v_i$.
If required, we can additionally ensure that all these fixed-point states are pairwise distinct by asserting the hard constraint $\delta_{s, t} \equiv \bigvee_{1 \leq i \leq n} s_i \neq t_i$ for each pair of fixed-point states $s \ne t$.

\begin{equation}
	\label{eq:fixed-point-macro}
	\sigma_{s} \equiv \bigwedge_{1 \leq i \leq n} s_i = f_i(s_{r_1^i}, \ldots, s_{r_k^i})
\end{equation}

\smallskip
\noindent
\emph{Observation constraints.}
As observation constraints consist of comparisons between state variables and constants, they naturally translate to the SMT representation.
%Specifically, constraints of the form $s_i^j \bowtie s_{i'}^{j'} \in \mathcal{C}$ %(where $\bowtie \in \{=, \neq, \geq, \leq, > , <\}$)
%are encoded as $\varsigma_i^j \bowtie \varsigma_{i'}^{j'}$, and those of the form $s_i^j \bowtie k \in \mathcal{C}$ as $\varsigma_i^j \bowtie k$.
Each $c \in \mathcal{C}_{\hard}$ is asserted as a hard constraint. For each level $1 \leq i \leq l$, each $c \in \mathcal{C}_{i}$ is asserted as a soft constraint with weight $w(c)$ and priority level $i$.

\medskip
Finally, for each variable or uninterpreted function with a multi-valued
domain, we also introduce a hard constraint asserting its domain
bounds. For Boolean variables, the encoding uses propositional
variables instead of bounded integers. The values of propositional
variables are naturally ordered by $\mathit{false} < \mathit{true}$,
which can be expressed in propositional logic as a simple implication.

%\subsection{Optimization}
%\label{subsec-optimization}
%Constraints, weights, priority classes

\enlargethispage*{5mm}
\section{Data-Informed Steady-State Constraints}\label{sec:constraints}

\begin{figure}[t]
	\centering
	\includegraphics[width=1.0\linewidth]{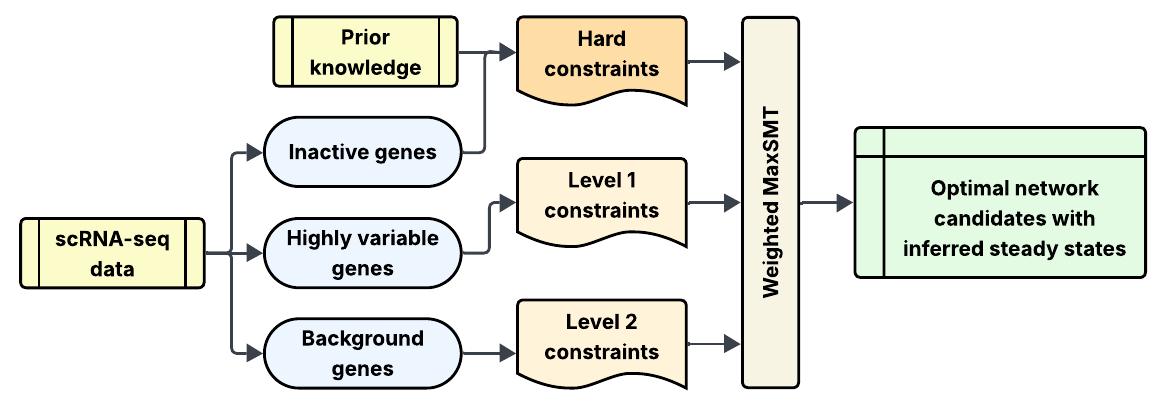}
	\caption{An overview of the data-informed workflow for MVN inference used for evaluation in this paper. Other applications of our weighted MaxSMT approach are free to introduce their own dedicated strategies to derive constraints from biological observations.}
	\label{fig:workflow}
\end{figure}

The method presented in Sections~\ref{sec:mvn-inference} and \ref{sec:mvn-inference-encoding} is general: it accommodates any combination of essentiality, monotonicity, and \emph{weighted} observation constraints. Ultimately, it is up to the modeller to select constraints that best reflect the biological reality. However, a practical MVN inference method should also provide guidance on how to construct these constraints. As such, we next discuss various approaches to deriving constraints from biological data and prior knowledge.

To ground the discussion in a specific modelling scenario, assume that the influence graph represents a gene regulatory network, and the fixed points correspond to different cell-type phenotypes. For each cell type, we have a collection of scRNA-seq expression data. The inferred model provides a plausible explanation for the differences in gene expression observed across these cell types. The specific approach we propose for this scenario is also illustrated in Fig.~\ref{fig:workflow}.

\enlargethispage*{4mm}
\paragraph{Prior-Knowledge Constraints}

A realistic modelling scenario (e.g.,~\cite{greene2025boolean}) is often subject to prior-knowledge assumptions that are believed to be true but are supported indirectly---such as through literature or curated databases---rather than by direct, quantifiable observations. For example, we may wish to incorporate cell-type marker genes compiled through a literature search, or to ensure the existence of a cell-type phenotype present in a pre-existing model.

Prior-knowledge constraints generally belong to the $\mathcal{C}_{\hard}$ level of observation constraints, as we expect them to be satisfied by any valid model. Nevertheless, if the prior-knowledge proves to be contradictory, we can instead introduce them as soft constraints within a dedicated level $\mathcal{C}_1$ using equal weights (e.g., $w(c) = 1$ for each $c \in \mathcal{C}_1$). This maximises the number of satisfied prior-knowledge constraints, allowing us to further reconsider the ones that are not satisfied.

\paragraph{Level-Based Observation Constraints}

Constraints on the biological steady states typically appear as \emph{binarised} or otherwise \emph{discretised} observations~\cite{chevalier2025data,herault2023novel,yordanov2016method}. These discrete observations are derived from real-world measurements, such as bulk~\cite{chagas2023boolean} or single-cell RNA~\cite{herault2023novel} gene expression, or other observation techniques (e.g., Western blots~\cite{singh2012boolean}, proteomics~\cite{mayer2016boolean}). These are then processed by a chosen discretisation tool, such as \texttt{scBoolSeq}~\cite{sc-bool-seq} or \texttt{BiTrinA}~\cite{mussel2016bitrina}. Various ad-hoc discretisation strategies or portfolios of tools are often employed as well~\cite{chevalier2025data,herault2023novel}.

In our framework, each discretised measurement results in an observation constraint $s_i^j = k$ (with $k \in D_i$). Some measurements are so reliable that we may introduce them as hard constraints. For example, we assert that $s_i^j = 0$ is a hard constraint for each gene $i$ that is never expressed within cell type $j$ (\emph{inactive genes} in Fig.~\ref{fig:workflow}). Nevertheless, most discretised observations are subject to some uncertainty, meaning we should treat them as soft constraints. Discretisation methods typically provide a confidence score reflecting the quality of the discretisation result, which we can use directly as the constraint weight $w(s_i^j = k)$. This ensures the inference process prioritises constraints obtained from high-quality discretisations. We have already encountered constraints of this type in Scenario I of Example~\ref{ex:running}.

%\enlargethispage*{4mm}
%\paragraph{Limitations of Level-Based Observation Constraints}

It is important to note, however, that these confidence scores may not have a clear statistical interpretation, and they may not be comparable between different tools or datasets. Even for interpretable weights, their interpretation can be subject to additional assumptions. For example, considering the constraint $s_i^j = 1$ with weight $\xi$, we could interpret $\xi$ as the estimated probability that the expression of gene $i$ in cell type $j$ exceeds a specific threshold $\theta$. However, this interpretation relies on an implicit assumption that the chosen $\theta$ accurately captures a qualitative switch in the underlying behaviour of gene $i$. Two discretisation methods may not agree on the same threshold, making their weights incomparable. Consequently, we do not recommend mixing constraint weights from diverse sources within one level $\mathcal{C}_i$ of observation constraints without careful consideration.

\paragraph{Ordering-Based Constraints}

To eliminate reliance on assumed discretisation thresholds, we can consider constraints based on ordering. For a gene $i$ and a pair of steady states $s^a$ and $s^b$, we introduce three mutually exclusive soft constraints: $s_i^a < s_i^b$, $s_i^a = s_i^b$, and $s_i^a > s_i^b$. The weights assigned to these soft constraints reflect our confidence in the ordering of the corresponding values. Depending on the nature of our measurements, we can again use different tools to derive these constraint weights. For example, given RNA expression, we can use the results of differential expression analysis~\cite{soneson2013comparison} as the basis for suitable weights. We have already encountered constraints of this type in Scenario II of Example~\ref{ex:running}.

As a concrete use case, let $\xi$ be the estimated probability that the expression of gene $i$ in cell type $a$ is lower than in cell type $b$. Then, we set $w(s_i^a < s_i^b) = \xi$, $w(s_i^a = s_i^b) = 1$, and $w(s_i^a > s_i^b) = 1 - \xi$. Compared to level-based constraints, this approach is more cautious: it reflects the ordering between the respective observations, but it does not require that a \emph{quantitative} difference in gene expression also results in a \emph{qualitative} difference within the inferred model. This is because the $s_i^a = s_i^b$ constraint is always the optimal choice, forcing the solver to introduce qualitative differences only in situations where they are required by other, higher-priority constraints (e.g., distinctness of fixed points, prior-knowledge, monotonicity, essentiality). In the absence of higher-priority constraints, an optimal model simply consists of constant update functions.

As an alternative, we also propose the weight assignment $w(s_i^a < s_i^b) = \xi^2$, $w(s_i^a = s_i^b) = 2 \xi (1-\xi)$, and $w(s_i^a > s_i^b) = (1 - \xi)^2$. Here, equality is only optimal if the estimated probability falls within the interval $(\frac{1}{3}, \frac{2}{3})$. This is conceptually similar to level-based discretisation, as it assumes that a quantitative change in gene expression should also result in a different discrete level. However, compared to the level-based approach, this formulation is more flexible in the case of multi-valued models: instead of fixed levels established by the discretisation tool, the solver can select a level assignment that optimises both the quantitative changes in gene expression and other dynamical properties of the model. In other words, beyond absolute gene expression, the solver can consider additional factors, such as the qualitative effect of the gene on its downstream regulation targets.

\paragraph{Priority of Ordering-Based Constraints}

In our tested workflow, we propose the following approach to using ordering-based constraints: For genes assumed to be significant contributors to the observed cell-type phenotypes, we use the second approach and place the constraints into the $\mathcal{C}_1$ level. These should primarily be known marker genes, but can also include other selected genes with highly variable expression (\emph{highly variable genes} in Fig.~\ref{fig:workflow}). Meanwhile, in the absence of further evidence or assumptions, we use the first, more cautious approach and place these constraints in the $\mathcal{C}_2$ level (\emph{background genes} in Fig.~\ref{fig:workflow}). In other words, we require that the inferred model primarily explains the differences in gene expression that we associate with the cell-type phenotypes, with a secondary requirement that any other genes included in the model follow the ordering derived from their RNA-seq expression. However, we should note that this approach still depends on the user's choice of phenotype-determining genes, the criteria for which are the responsibility of the model authors. We view this as a somewhat unavoidable aspect of qualitative modelling, where part of the model design is to determine which qualitative differences are to be explained by the model.

\section{Experiments}

To evaluate our method, we perform a thorough investigation of its capabilities on 502 benchmark instances derived from a realistic modelling scenario based on scRNA-seq data and a prior-knowledge gene regulatory network. All code, data, and computed results are available as a Zenodo archive.\footnote{\url{https://doi.org/10.5281/zenodo.19508672}}

\paragraph{Implementation}

Our MaxSMT encoding is implemented in a Rust-based open-source tool available on GitHub.\footnote{\url{https://github.com/sybila/biodivine-algo-smt-inference}} As input, the tool accepts an influence graph in the \texttt{.aeon} format~\cite{benevs2022aeon}, with additional annotations describing the number of fixed points as well as the hard and soft constraints to be applied during model inference. The tool then constructs the SMT query as proposed in Section~\ref{sec:mvn-inference-encoding} and delegates the optimisation to the Z3 SMT solver~\cite{de2008z3}.

\subsection{Test instances}

To prepare a challenging environment for our method, we have constructed a collection of large-scale inference problems, following the workflow outlined in Section~\ref{sec:constraints} and Fig.~\ref{fig:workflow}. However, we emphasise that in this paper, we focus on evaluating the technical aspects of the inference method, rather than on the biological validation of the models inferred during the evaluation.

\paragraph{Modelling scenario}

\begin{table}[t]
	\caption{Summary of the tested inference instances. Instances are grouped by the number of included cell types. For each group, the table lists the total number of test instances, along with the ranges of genes, regulations, and soft constraints within that group. The final two columns report the number of instances where an optimal solution was successfully found within a six-hour time limit for either the first soft constraint level ($\mathcal{C}_1$) or for both levels simultaneously $(\mathcal{C}_1, \mathcal{C}_2)$. Bold values indicate that 100\% of the instances within that group were solved to optimality.}
	\label{tab:evaluation-overview}
	\centering
	\setlength{\tabcolsep}{4pt}
	\begin{tabular}{@{}rrrrrrr@{}}
		\toprule
		\multirow{2}{28pt}{\raggedleft \textbf{Cell Types}} & \multirow{2}{*}{\textbf{Instances}} & \multirow{2}{*}{\textbf{Genes}} & \multirow{2}{*}{\textbf{Regulations}} & \multirow{2}{64pt}{\raggedleft \textbf{Soft Constraints}} & \multicolumn{2}{c}{\textbf{Fully Solved}} \\
		\cmidrule(l){6-7}
		& & & & & $\mathcal{C}_1$ & $(\mathcal{C}_1, \mathcal{C}_2)$ \\
		\midrule
		2 & 36 & 183--612 & 882--2,434 & 362--869 & \textbf{36} & \textbf{36} \\
		3 & 84 & 211--865 & 1,037--3,182 & 1,255--4,326 & \textbf{84} & \textbf{84} \\
		4 & 126 & 280--1,004 & 1,269--3,666 & 3,278--10,676 & \textbf{126} & 89 \\
		5 & 126 & 349--1,136 & 1,526--4,080 & 6,743--20,672 & \textbf{126} & 28 \\
		6 & 84 & 409--1,225 & 1,728--4,316 & 11,847--34,158 & \textbf{84} & 8 \\
		7 & 36 & 524--1,275 & 2,027--4,480 & 21,110--50,249 & 32 & 0 \\
		8 & 9 & 955--1,307 & 3,445--4,599 & 51,108--69,238 & 2 & 0 \\
		9 & 1 & 1,326 & 4,664 & 90,964  & 0 & 0 \\
		\bottomrule
	\end{tabular}
\end{table}

As the basis for our test instances, we consider neural differentiation in the mouse cerebral cortex, as mapped by the single-cell RNA atlas from~\cite{DiBella2021} and a prior-knowledge gene regulatory network available in the Omnipath database~\cite{omnipath}. Here, we briefly describe the process used to create the inference constraints from these inputs. Additional technical details are given in Appendix~\ref{appendix:data-preparation} of~\cite{huvar2026inference} and the reproducibility artefact. After initial filtering, we select nine cell types that correspond to various fully differentiated neural cells to act as phenotypes. We then consider all 502 possible unique combinations of two or more cell types when constructing individual test instances.

For each test instance, we first construct a prior-knowledge network. We exclude any genes that are not covered by Omnipath and any genes that are inactive in all considered cell types. Consequently, the size of the influence graph is different for each test instance, as each cell type introduces different active genes with available prior-knowledge regulations. Overall, the influence graph size ranges from 183 to 1,326 genes and from 882 to 4,664 regulations. Additional information about the test instances is available in Table~\ref{tab:evaluation-overview}. We primarily report results for Boolean domains, as Boolean model inference is more common in literature. In Appendix~\ref{appendix:additional-results} of~\cite{huvar2026inference}, we also compare the runtime with multi-valued instances where the gene domain size is the number of its outgoing regulations. Overall, we find that using multi-valued domains does introduce a non-trivial, order of magnitude slowdown, but is still tractable for many test instances in our benchmark. Furthermore, our test represents a worst-case scenario; in practical applications, variable domains are often much smaller.

\paragraph{Constraint preparation}

First, if a gene is not expressed in a specific cell type at all, we introduce this observation as a hard constraint. Then, to derive the ordering-based constraints, we train an \texttt{scvi-tools}~\cite{scvi2022python} statistical model on the scRNA-seq data, assuming a zero-inflated negative binomial distribution of gene expression. In other words, the true expression of each gene is modelled using a negative binomial distribution that is subject to stochastic dropout, explaining the observed zero-inflation in lowly expressed genes (see also Appendix~\ref{appendix:data-preparation} of~\cite{huvar2026inference}). We then use this learned distribution to estimate for each gene $i$ the probability $\xi$ that its expression is lower in cell type $a$ compared to cell type $b$. Finally, for each pair of cell types, we identify candidate marker genes as those where the difference in expression is not only highly likely ($\xi > 0.8$), but also substantial in magnitude (at least 80\% of the gene's maximal expression across all cell types). As discussed in Section~\ref{sec:constraints}, we use the second type of weight assignment (prioritising inequalities) for the identified marker genes to construct soft constraints $\mathcal{C}_1$, and the first type of weight assignment (prioritising equalities) for the remaining active genes to construct $\mathcal{C}_2$. This yields a complete inference problem specification that can then be used for evaluation.

\subsection{Evaluation results}

\begin{figure}[t]
	\centering
	\includegraphics[width=0.48\linewidth]{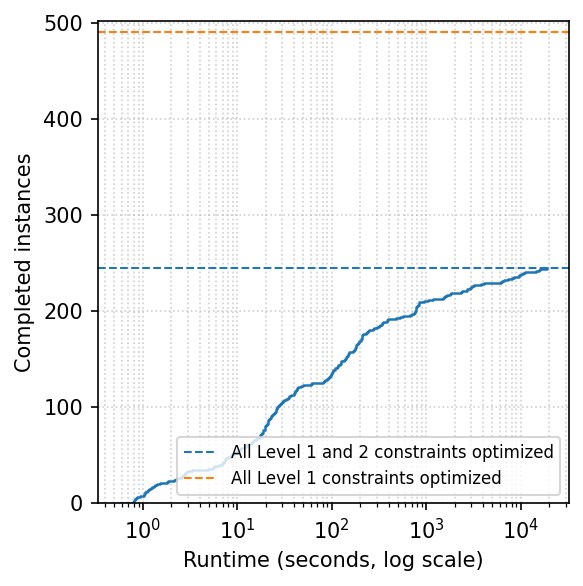}
	\includegraphics[width=0.48\linewidth]{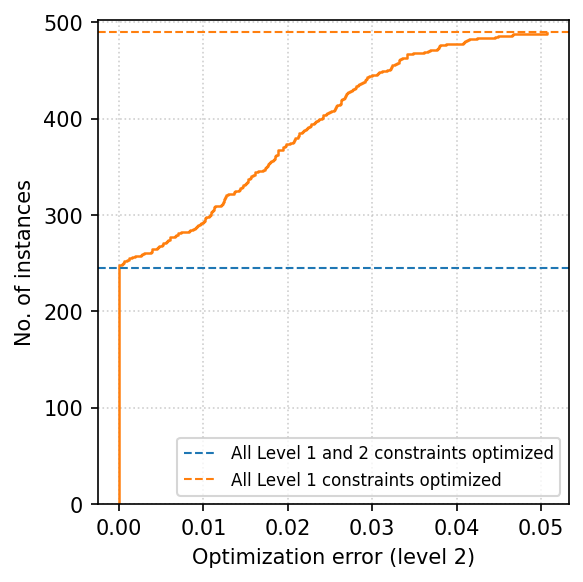}
	\caption{Cumulative plots showing the performance of our method on the tested benchmark instances. Left: The 245/502 instances where an optimal solution was found within the 6h timeout. The plot shows the number of instances (y-axis) completed within a certain time (x-axis; logarithmic). Right: The 490/502 instances where an optimal solution for $\mathcal{C}_1$ constraints was found within the 6h timeout. The plot shows the number of instances (y-axis) with a less than or equal relative optimisation error in $\mathcal{C}_2$ constraints (x-axis). The relative optimisation error is the size of the confidence interval produced by the MaxSMT solver relative to the sum of constraint weights in $\mathcal{C}_2$.}
	\label{fig:performance}
\end{figure}

Experiments were conducted with a six-hour timeout on a Ryzen 9900X3D CPU with 128GB RAM, running up to 10 instances in parallel and a 6h time limit. We rely on the fact that in the absence of an optimal solution, the solver still reports the best solution discovered before the timeout and bounds on the optimal weight within the first unsolved level of constraints. In Table~\ref{tab:evaluation-overview}, we report the number of instances where (a) the solver successfully optimised constraint level $\mathcal{C}_1$; (b) the solver successfully optimised both $\mathcal{C}_1$ and $\mathcal{C}_2$, producing a fully optimal solution. Here, we see that fully optimal solutions are found in instances with up to 6 cell types, and optimal solutions w.r.t. $\mathcal{C}_1$ are found for up to 8 cell types.

In Fig.~\ref{fig:performance} (left), we summarise the runtime of 245 problem instances that were fully solved. Meanwhile, for the 490 instances where at least $\mathcal{C}_1$ was fully optimised, we report the normalised optimisation error in Fig.~\ref{fig:performance} (right). This error is computed as the size of the interval that must contain all optimal solutions, which is reported by the solver, relative to the sum of constraint weights in $\mathcal{C}_2$. Overall, we see that for all 490 instances, the solver found a solution within 5\,\% of the total weight. Detailed visualisation with respect to inference problem size and comparison with multi-valued results is then given in Appendix~\ref{appendix:additional-results} of~\cite{huvar2026inference}. Overall, we show that our method can handle even very large instances ($\geq$1,200 genes, 35,000 constraints), and in the absence of an optimal solution provides results that are close to optimal (within 5\% of the total weight).

\enlargethispage*{4mm}
\section{Discussion}\label{sec:discussion}

In this work, we present a novel method for Boolean and multi-valued network inference from uncertain experimental data based on weighted MaxSMT. The novelty of the approach lies in the fact that it
\begin{enumerate*}[(a)]
	\item takes into account the grade of uncertainty in the experimental observations, encoding it using rational weights, and
	\item allows us to express more general comparison constraints instead of just relying on a binarisation (or discretisation) method to preprocess the data.
\end{enumerate*}

Experimental results show that the method is applicable to real-world data and scales well to deal with a large number of network variables. Moreover, our approach admits an \emph{anytime} implementation, provided the SMT optimiser supports it. Hence, even for larger instances, the computation can be stopped at any time while still returning a feasible solution together with a bound on its suboptimality, i.e., how close its weight is to the optimal one.

%Currently, the method only focuses on fixed-point states. This has some advantages, as it makes the approach applicable regardless of the concrete state-transition semantics. However, in the future, we plan to expand the method to include more complex long-term behaviour phenomena, such as trap spaces.

Note that we currently only use a specific fragment of the possibilities offered by the SMT encoding. In the future, we plan to include objective functions that allow, e.g., minimising the number of regulations, minimising the domain size (while maintaining the optimal weight), and other forms of optimisation. We also plan to tackle complex long-term behaviour phenomena, such as trap spaces.

Moreover, we plan to provide a more efficient enumeration of the resulting models. Currently, enumeration is possible by instructing the underlying SMT solver to avoid the previously generated solutions and then calling the procedure again. In this way, we get the candidate models one by one. A~more efficient approach, planned for the future, would produce candidate models in larger batches, e.g., by describing them in terms of partially-specified functions.

A theoretical limitation of our approach is that we cannot require the model to have \emph{exactly} the prescribed number of fixed points. Encoding this would require alternating quantifiers, which makes the problem significantly harder to solve.

\subsubsection*{Acknowledgements}
The work was supported by the MUNI/JS/1954/2025 project of Masaryk University and the GA26-21507S project of the Czech Science Foundation.

\subsubsection*{Competing interests}
The authors have no competing interests to declare that are relevant to the content of this article.

\bibliographystyle{splncs04}
\bibliography{references}

\clearpage
\appendix
\section{Example – More Details}\label{app:example}

There are $393\,030$ optimal solutions for Scenario I of
Example~\ref{ex:running}, all of which have the same valuation of the fixed-point states, namely (0, 2, 1), (1, 2, 1); the weight of the optimal solutions is 3.59. One of the possibilities
is given by the following update functions ($a$, $b$, $c$ stand for the values of variables $A$, $B$, $C$, respectively):
\begin{align*}
	F_A(a, b, c) &=
	\begin{cases}
		1 & a = 1 \land (b \le 1 \lor c \ge 1) \\
		0 & \text{otherwise}
	\end{cases}
	\\
	F_B(a, b, c) &=
	\begin{cases}
		2 & a = 1 \lor c \ge 1 \\
		1 & \text{otherwise}
	\end{cases}
	\\
	F_C(a, b, c) &=
	\begin{cases}
		1 & b = 2 \\
		0 & \text{otherwise}
	\end{cases}
\end{align*}
Note that this solution has one additional fixed-point state, namely (0, 1, 0)---as mentioned in Section~\ref{sec:discussion}, our method cannot currently enforce that the specified fixed-point states are the only ones in the resulting model.

In the case of Scenario II without the distinctness requirement, the possible values for the fixed points are the following:
\begin{itemize}
	\item (1, 1, 1), (1, 1, 1), (1, 0, 0)
	\item (1, 1, 2), (1, 1, 2), (1, 0, 0)
	\item (1, 1, 2), (1, 1, 2), (1, 0, 1)
	\item (1, 2, 1), (1, 2, 1), (1, 0, 0)
	\item (1, 2, 1), (1, 2, 1), (1, 1, 0)
	\item (1, 2, 2), (1, 2, 2), (1, 0, 0)
	\item (1, 2, 2), (1, 2, 2), (1, 0, 1)
	\item (1, 2, 2), (1, 2, 2), (1, 1, 0)
	\item (1, 2, 2), (1, 2, 2), (1, 1, 1)
\end{itemize}
The total number of optimal solutions is 6\,603\,477, and their weight is 1.50.
One of the possibilities (with fixed points (1, 2, 1), (1, 2, 1), (1, 1, 0)) is given by the following update functions:
\begin{align*}
	F_A(a, b, c) &=
	\begin{cases}
		1 & a = 1 \lor b \le 1 \lor c = 2 \\
		0 & \text{otherwise}
	\end{cases}
	\\
	F_B(a, b, c) &=
	\begin{cases}
		2 & c \ge 1 \\
		1 & a = 1 \land c = 0 \\
		0 & \text{otherwise}
	\end{cases}
	\\
	F_C(a, b, c) &=
	\begin{cases}
		1 & b = 2 \\
		0 & \text{otherwise}
	\end{cases}
\end{align*}

If we add the distinctness requirement to Scenario II, there are $130\,764$ optimal solutions, all of which have the same valuation of the fixed-point states, namely (1, 2, 2), (1, 1, 1), (1, 0, 0).
The weight of these optimal solutions is 1.45.
One of the possibilities is given by:
\begin{align*}
	F_A(a, b, c) &=
	\begin{cases}
		1 & b = 0 \lor (a = 1 \land c \ge 1) \\
		0 & \text{otherwise}
	\end{cases}
	\\
	F_B(a, b, c) &=
	\begin{cases}
		2 & a = 1 \land c = 2 \\
		1 & (a = 1 \land c = 1) \lor (a = 0 \land c = 2)\\
		0 & \text{otherwise}
	\end{cases}
	\\
	F_C(a, b, c) &= b
\end{align*}

\section{Evaluation -- Data Preparation}
\label{appendix:data-preparation}

\begin{table}[]
	\caption{Summary of the labelled cell types in the neural differentiation RNA dataset, reporting whether the cell type is fully differentiated (terminal), its type, number of cells in the cell type, as well as cell-wise and gene-wise mean absolute error produced by the learned \texttt{scvi-tools} statistical model.}
	\centering
	\begin{tabular}{l|c|c|c|c|c}
		Cell type & terminal & type & cell count & cell-MAE & gene-MAE \\ \midrule
		Apical progenitors          & no  & neural   & 14668 & 0.19 & 7.21 \\
		UL CPN (Layer 2\&3)         & yes & neural   & 13401 & 0.38 & 6.70 \\
		Migrating neurons           & no  & neural   &  9937 & 0.55 & 6.79 \\
		Interneurons                & yes & neural   &  7579 & 0.59 & 4.95 \\
		Intermediate prog.          & no  & neural   &  7002 & 0.33 & 5.65 \\
		Layer 4                     & yes & neural   &  5247 & 0.36 & 4.99 \\
		CThPN                       & yes & neural   &  4424 & 0.64 & 4.44 \\
		DL CPN (Layer 5\&6)         & yes & neural   &  3043 & 0.42 & 3.99 \\
		SCPN                        & yes & neural   &  2764 & 0.52 & 3.70 \\
		Astrocytes                  & yes & glial    &  2716 & 0.86 & 3.64 \\
		Immature neurons            & no  & neural   &  2692 & 0.55 & 4.12 \\
		Oligodendrocytes            & yes & glial    &  1005 & 0.82 & 2.79 \\
		Cajal Retzius cells         & yes & neural   &   437 & 0.76 & 1.90 \\
		Near projecting (NP)        & yes & neural   &   385 & 0.46 & 2.23 \\
		Red blood cells             & yes & vascular &   294 & \textbf{5.57} & 3.18 \\
		VLMC                        & yes & vascular &   223 & 0.29 & 1.83 \\
		Pericytes                   & yes & vascular &   219 & 0.47 & 1.71 \\
		Endothelial cells           & yes & vascular &   214 & 1.56 & 1.62 \\
		Layer 6b                    & yes & neural   &   193 & 0.37 & 1.61 \\
		Microglia                   & yes & glial    &   187 & 0.87 & 1.87 \\
	\end{tabular}
	\label{tab:cell-type-summary}
\end{table}

\paragraph{Neural differentiation dataset}
We consider an scRNA-seq dataset covering neural differentiation in mouse cerebral cortex~\cite{DiBella2021}. The published dataset includes 80\,467 cells with known cell type annotations (not all cells in the raw data are annotated; we do not consider these) and 27\,933 genes. Out of these annotated cells, we select 76\,630 cells by removing those designated as low quality in the original paper, or lacking clear cell type identity (e.g., doublets). This results in 20 neural, glial, and vascular cell types as summarised in Table~\ref{tab:cell-type-summary}. Subsequently, we also select 17\,920 genes that are in at least one cell type expressed by $\geq3$ cells per 1\,000. The remaining genes are assumed to be not-expressed and are not considered in the modelling process. Note that this is a relatively conservative criterion; other studies often consider only a much smaller subset of highly variable genes. Here, our intention is to preserve low-expression genes that may be important but are otherwise hard to observe.

\paragraph{Learning gene expression ordering using \texttt{scvi-tools}}
We use the filtered dataset to train an \texttt{scvi-tools}~\cite{scvi2022python} statistical model, assuming a zero-inflated negative binomial distribution of gene expression. In other words, the true expression of each gene is modelled using a negative binomial distribution that is subject to stochastic dropout, causing zero-inflation in genes with low expression. To train the model, we configure \texttt{scvi-tools} with three hidden layers using 128 nodes, 30 latent dimensions, 0.1 dropout rate, and 200 training epochs.

This is slightly elevated compared to the default settings, but is often recommended for datasets covering a wide range of cell types~\cite{scvihub2025scvi}. To assess the model quality and the robustness of the learning process, we consider four configurations of hyper-parameters that we compare in depth in the reproducibility artefact. While our chosen configuration achieves the best performance, the absolute differences between the learned models point towards a robust learning outcome.

To evaluate the quality of individual models, we use the recommended posterior predictive check feature of the \texttt{scvi-criticism}~\cite{scvihub2025scvi} module to assess the cell-wise and gene-wise coefficient of variation (i.e., the ratio of standard deviation and mean). All four trained models achieve $R^2$ scores well above the $0.4$ threshold recommended by the tool authors. Our chosen model achieves an $R^2$ score of 0.86 (cell-wise) and 0.91 (gene-wise), with the mean absolute error (MAE) of 0.48 (cell-wise) and 2.36 (gene-wise). These match or exceed previous high-quality statistical models constructed by the authors of scVI~\cite{tabula2022tabula}. We also compute the mean absolute error for each cell type separately to assess whether the class imbalance in our dataset (some cell types are significantly more abundant) affects the performance of scVI. These are also shown in Table~\ref{tab:cell-type-summary}. As we can see, most cell types are represented fairly well (they are close to the mean absolute error of the whole dataset); the only outlier are red blood cells. Note that the gene-wise MAE is in general expected to be higher than cell-wise MAE, because genes with rare, spiky expression typically have a very high coefficient of variation.

\newpage

\section{Evaluation -- Additional Results}
\label{appendix:additional-results}

\begin{figure}
	\centering
	\includegraphics[width=1.0\linewidth]{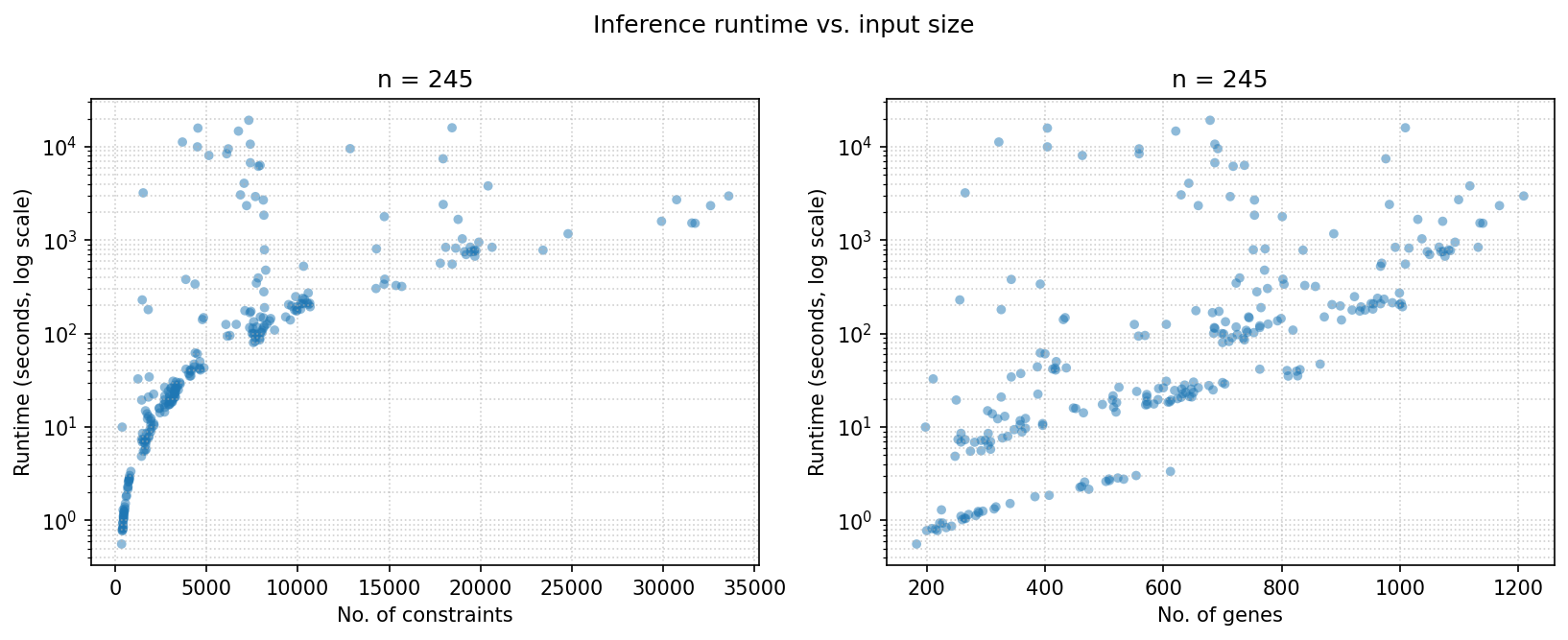}
	\caption{Detailed analysis of runtime with respect to specification size (the number of soft constraints) and influence graph size (the number of genes).}
	\label{fig:runtime-vs-specification}
\end{figure}

\begin{figure}
	\centering
	\includegraphics[width=1.0\linewidth]{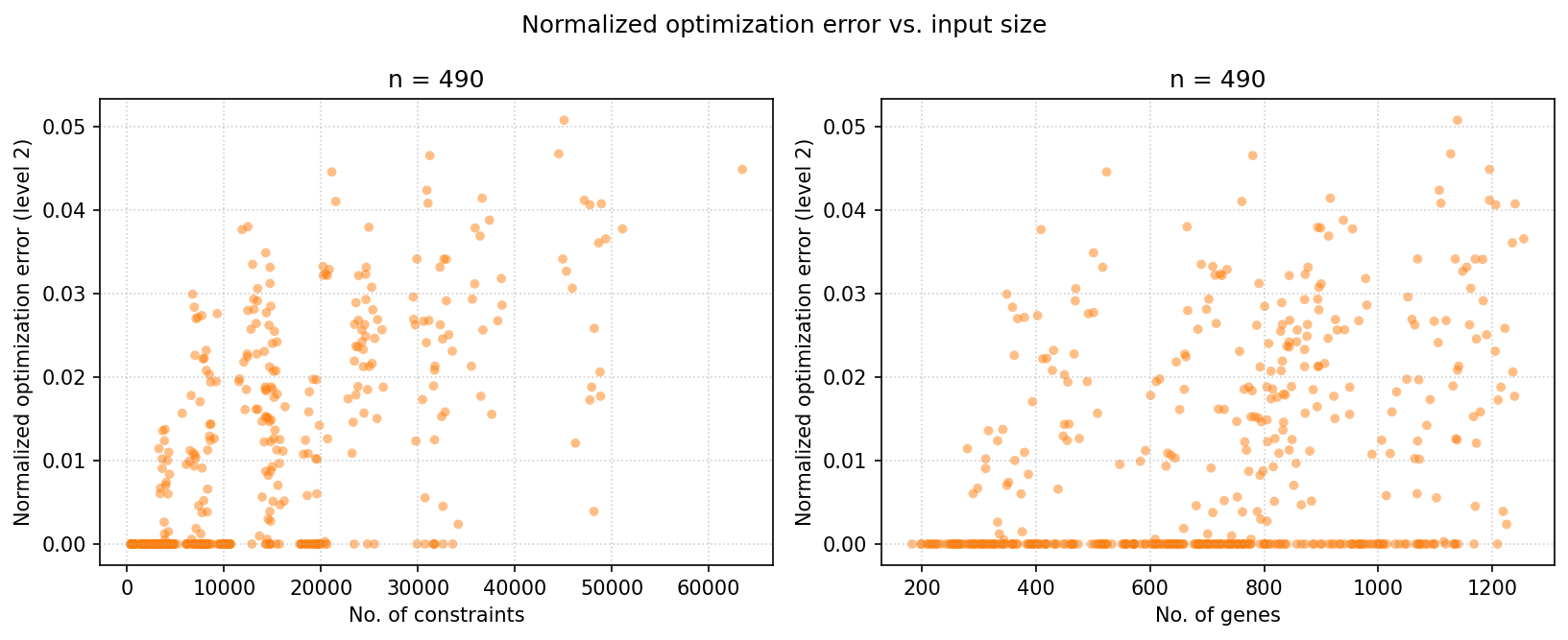}
	\caption{Detailed analysis of the normalised optimisation error (size of the error interval relative to the sum of all weights) for $\mathcal{C}_2$ constraints with respect to specification size (the number of soft constraints) and influence graph size (the number of genes).}
	\label{fig:error-vs-specification}
\end{figure}

\begin{figure}
	\centering
	\includegraphics[width=0.5\linewidth]{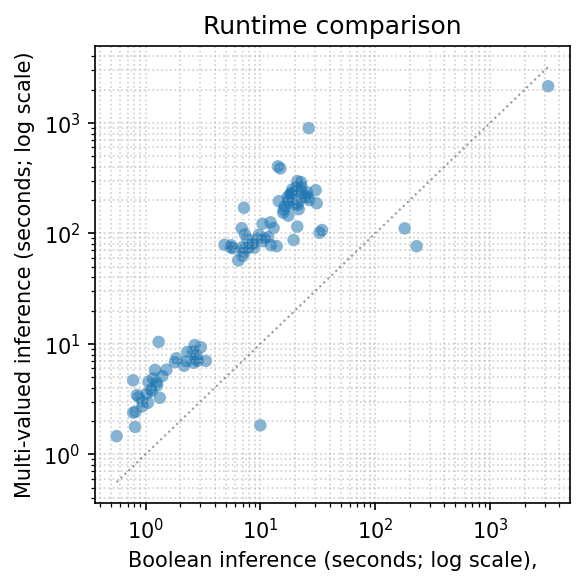}
	\caption{Runtime comparison of the first 120 problem instances (using two and three cell types) between Boolean and multi-valued problem formulations.}
	\label{fig:runtime-comparison-scatter}
\end{figure}

\end{document}